\documentclass[prb,preprintnumbers,superscriptaddress,amsmath,amssymb,amsart,twocolumn]{revtex4}

\usepackage{graphicx}
\usepackage{dcolumn}
\usepackage{bm}
\usepackage{latexsym,epsfig}
\usepackage{chemarrow}
\usepackage{xcolor}
\usepackage{ulem}
\begin{document}

\title{Oxygen on-site Coulomb energy in Pr$_{1.3-x}$La$_{0.7}$Ce$_x$CuO$_{4}$ and  Bi$_2$Sr$_2$CaCu$_2$O$_{8+\delta}$ and its relation with Heisenberg exchange} 

\author{A. Chainani}
\affiliation{Condensed Matter Physics Group, National Synchrotron Radiation Research Center, Hsinchu 30076, Taiwan}
\affiliation{RIKEN SPring-8 Centre, 1-1-1 Sayo-cho, Hyogo 679-5148, Japan}
\author{M. Horio}
\affiliation{Department of Physics, The University of Tokyo, 7-3-1 Hongo, Bunkyo-ku, Tokyo 113-0033, Japan}
\affiliation{Institute for Solid State Physics, The University of Tokyo, Kashiwa, Chiba 277-8581, Japan}
\author{C.-M. Cheng}
\affiliation{Condensed Matter Physics Group, National Synchrotron Radiation Research Center, Hsinchu 30076, Taiwan}
\author{D. Malterre}
\affiliation{Institut Jean Lamour, Universit\'{e} de Lorraine, UMR 7198 CNRS, BP70239, 54506 Vandoeuvre l\'{e}s Nancy, France}
\author{K. Sheshadri} 
\affiliation{226, Bagalur, Bangalore North, Karnataka State, India 562149}
\author{M. Kobayashi}
\affiliation{Department of Electrical Engineering and Information Systems and Center for Spintronics Research Network, The University of Tokyo, 7-3-1 Hongo, Bunkyo-ku, Tokyo 113-8656, Japan}
\author{K. Horiba}
\affiliation{National Institutes for Quantum and Radiological Science and Technology (QST), Sayo, Hyogo 679-5148,Japan.}
\author{H. Kumigashira}
\affiliation{Institute of Multidisciplinary Research for Advanced Materials (IMRAM), Tohoku University, Sendai 980-8577, Japan.}
\author{T. Mizokawa}
\affiliation{Department of Applied Physics, Waseda University, Shinjuku, Tokyo 169-8555, Japan.}
\author{M. Oura}
\affiliation{RIKEN SPring-8 Centre, 1-1-1 Sayo-cho, Hyogo 679-5148, Japan}
\author{M. Taguchi}
\thanks{Present address: Toshiba Nanoanalysis Corporation, Kawasaki 212-8583, Japan}
\affiliation{RIKEN SPring-8 Centre, 1-1-1 Sayo-cho, Hyogo 679-5148, Japan}
\author{Y. Mori}
\affiliation{Department of Applied Physics, Tohoku University, Sendai 980-8579, Japan.}
 \author{A. Takahashi} 
 \affiliation{Department of Applied Physics, Tohoku University, Sendai 980-8579, Japan.}
 \author{T. Konno} 
 \affiliation{Department of Applied Physics, Tohoku University, Sendai 980-8579, Japan.}
 \author{T. Ohgi} 
 \affiliation{Department of Applied Physics, Tohoku University, Sendai 980-8579, Japan.}
 \author{H. Sato} 
 \affiliation{Department of Applied Physics, Tohoku University, Sendai 980-8579, Japan.}
 \author{T. Adachi}
 \affiliation{Department of Engineering and Applied Sciences, Sophia University, Tokyo 102-8554, Japan}
 \author{Y. Koike}
\affiliation{Department of Applied Physics, Tohoku University, Sendai 980-8579, Japan.}
\author{T. Mochiku}
\affiliation{National Institute for Materials Science, Tsukuba, Ibaraki 305-0047, Japan}
\author{K. Hirata}
\affiliation{National Institute for Materials Science, Tsukuba, Ibaraki 305-0047, Japan}
\author{S. Shin}
\thanks{deceased}
\affiliation{RIKEN SPring-8 Centre, 1-1-1 Sayo-cho, Hyogo 679-5148, Japan}
\author{M. K. Wu}
\affiliation{Institute of Physics, Academia Sinica, Taiwan}
\author{A. Fujimori}
\affiliation{Department of Physics, The University of Tokyo, 7-3-1 Hongo, Bunkyo-ku, Tokyo 113-0033, Japan}
\affiliation{Center for Quantum Science and Technology, and Department of Physics, National Tsing Hua University, Hsinchu 30013, Taiwan}
\affiliation{Condensed Matter Physics Group, National Synchrotron Radiation Research Center, Hsinchu 30076, Taiwan}

\date{\today}

\begin{abstract}

We study the electronic structure of electron-doped Pr$_{1.3-x}$La$_{0.7}$Ce$_{x}$CuO$_{4}$   (PLCCO ; $T_{c}$ = 27 K, x = 0.1) and hole-doped Bi$_2$Sr$_2$CaCu$_2$O$_{8+\delta}$  (Bi2212 ; $T_{c}$ = 90 K) cuprate superconductors using x-ray absorption spectroscopy (XAS) and resonant photoemission  spectroscopy (Res-PES). From Res-PES across the O K-edge and Cu L-edge, we identify the O 2p and Cu 3d partial density of states (PDOS) and their correlation satellites which originate in two-hole Auger final states. Using the Cini-Sawatzky method, analysis of the experimental O 2p PDOS shows an oxygen on-site Coulomb energy for PLCCO to be $U_{p}$ = 3.3$\pm$0.5 eV and for Bi2212, $U_{p}$ = 5.6$\pm$0.5 eV, while the copper on-site Coulomb correlation energy, $U_{d}$ = 6.5$\pm$0.5 eV for Bi2212. 
The expression for the Heisenberg exchange interaction $J$ in terms of the electronic parameters $U_{d}$, $U_{p}$, charge-transfer energy $\Delta$
and Cu-O hopping $t_{pd}$ obtained from a simple Cu$_2$O cluster model is used to carry out an optimization analysis consistent with $J$ known from scattering experiments. The analysis also provides the effective one band on-site Coulomb correlation energy $\tilde{U}$ and the effective hopping $\tilde{t}$. PLCCO and Bi2212 are shown to exhibit very similar values of $\tilde{U}$/$\tilde{t}$ $\sim$9-10, confirming the strongly correlated nature of the singlet ground state in the effective one-band model for both the materials.

\end{abstract}


\maketitle

\section{Introduction}
\label{introduce}

Since its discovery more than 35 years ago\cite{Bed}, an understanding of superconductivity in the high-transition temperature ($T_{c}$) cuprate superconductors continues to attract researchers even today.
Extensive experimental and theoretical efforts to understand the cuprates have identified important aspects of 
their electronic structure, such as spin- and charge-ordering,\cite{ZG,Machida,SWC,JT,Salkola,Wu,GG,JC,MT,Tabis,HJ} a d$_{x^2-y^2}$ type superconducting gap,\cite{Shen,Sato} role of anti-ferromagnetic correlations\cite{Julien,Lee,Chan},  electron-phonon coupling\cite{Lanzara}, a temperature and momentum-dependent pseudogap,\cite{Marshall,Ding} etc.  The charge ordering favours localization of carriers and competes with superconductivity of doped carriers in the CuO${_2}$ layers, thereby leading to novel  transport, thermodynamic, and spectroscopic phenomena which suggest quantum critical behavior \cite{Valla,Randeria,Pepin,Jacobs,Proust}. However, the origin for the high-$T_{c}$ superconductivity in the cuprates still remains an open problem.\cite{Keimer}

Several important models have emphasized the complex nature of the superconductivity and electronic structure of the cuprates. Starting with the one-band Hubbard model \cite{Anderson,Zhang}, theoretical models evolved along several different routes such as the  resonating valence bond theory \cite{Baskaran}, the three-band Hubbard model \cite{Varma01, Emery}, the $t-J$ model \cite{Spalek}, spin fluctuation theory  \cite{Schmalian}, marginal Fermi liquid theory \cite{Varma02}, pair density wave model \cite{Agterberg}, electron-phonon coupling-induced pairing which go beyond the BCS model \cite{Bishop}, etc. Although the origin of superconductivity  in the cuprates remains a challenge,
it is generally accepted that the quasi 2-dimensionality of the CuO$_{2}$ layers and strong on-site Coulomb correlations provide a suitable starting point for describing the electronic structure of the cuprates \cite{Anderson, Zhang, Baskaran, Varma01,Emery,Spalek, Schmalian, Varma02,Agterberg,Norman, arpesRMP,MottRMP,Das,Weber,Werner}.

Early studies using the Cini-Sawatzky method based on the two-hole Auger correlation satellite\cite{Cini,Sawatzky} showed that the O on-site Coulomb energy $U_{p}$ can be large ($\sim$5-6 eV) and close to the copper on-site Coulomb energy $U_{d}$ ($\sim$ 6-8 eV) in 
YBa$_{2}$Cu$_{3}$O$_{7}$ (YBCO)\cite{Marel,Balzarotti},  Bi$_2$Sr$_2$CaCu$_2$O$_{8+\delta}$ (Bi2212)\cite{Tjeng}, and La$_{2-x}$A$_{x}$CuO$_{4}$(A = Sr, Ba)\cite{BarDeroma,Fujimori,ZXS}. Further, $U_{p}$ $\sim$ $U_{d}$ is also known for several oxides across the 3d transition metal(TM) series : titanium/vanadium oxides(SrTiO$_{3}$, V$_{2}$O$_{3}$, VO$_{2}$, V$_{2}$O$_{5}$)\cite{Ishida,Post,Park}, LaMO$_{3}$ (M = Mn-Ni) perovskites\cite{AC,DD}, and cuprates(including Cu$_{2}$O and CuO) \cite{Marel,Balzarotti,Ghijsen,Tjeng,BarDeroma}. A theoretical study on rare-earth nickelates (RNiO$_{3}$) with values of $U_{d}$ (= 7 eV) and $U_{p}$ (= 5 eV) showed the relation of a novel charge-order involving ligand holes with the metal-insulator transition in RNiO$_{3}$.\cite{Johnston1} Very recently, the relation of the inter-site Heisenberg exchange interaction $J$ with $U_{d}$ and $U_{p}$ was recognized for the parent cuprates as well as hole doped cuprates\cite{Shesha}. In particular, it was shown that $J$ could be used as a bridge to connect the electronic parameters $\tilde{U}$ and  $\tilde{t}$ of the widely used effective one-band Hubbard model with the parameters $U_{d}$, $U_{p}$, $\Delta$ and $t_{pd}$ known from the three-band Hubbard model, cluster model calculations applied to core-level spectroscopy as well as resonant inelastic x-ray scattering,\cite{Shesha} and from {\it ab initio} electronic structure calculations\cite{Hirayama}. 

Surprisingly, there is no experimental estimate of $U_{p}$ using the Cini-Sawatzky method in electron-doped cuprates which possess CuO$_2$ planes without the apical oxygen site, i.e. the cuprates crystallizing in the so-called T' structure. 
For Bi2212, the estimate of $U_{d}$ and $U_{p}$ was made using known cluster model parameters\cite{Ghijsen} to explain the Res-PES spectra\cite{Tjeng}. While optimally doped Bi2212 (T$_C$ $\sim$ 90 K) has been extremely well-studied using soft and hard x-ray photoemission\cite{Eisaki,Tjeng,Bianconi,Nucker2,Brookes2001,Taguchi}, as well as low energy ARPES studies of its band dispersions and Fermi surfaces\cite{Shen,Lanzara,Marshall,Ding,Valla,arpesRMP}, there is no estimate of $U_{d}$ and $U_{p}$ using the experimental Cu $3d$ and O $2p$ PDOS. Thus, we felt it important to experimentally quantify on-site Coulomb energies in an electron-doped system in comparison with a well-studied hole-doped system. 
Further, recent studies on the T' structure PLCCO showed the importance of reduction annealing to achieve electron-doped superconductivity\cite{Horio1,Horio2,Horio_Thesis,Adachi}. From careful ARPES studies, it was shown that the superconducting state was found to extend over a wide electron doping range with an optimal T$_C$ $\sim$ 27 K\cite{CLin}. Interestingly, a sharp quasiparticle feature was observed on the entire Fermi surface 
of optimally-doped PLCCO with no signature of the antiferromagnetic (AF) pseudogap which indicated a reduced AF correlation length\cite{Horio1}. However, the superconducting gap still showed a d$_{x^2 - y^2}$ symmetry like the well-known results for the hole-doped Bi2212 \cite{Shen} and for electron-doped NCCO \cite{Sato}, and suggests the importance of spin-fluctuations as a viable source of pairing even for PLCCO\cite{Horio2}. 

In this work, we have used the Cini-Sawatzky method to obtain $U_{d}$ (= 6.5$\pm$0.5 eV for Bi2212) and $U_{p}$ values
(= 5.6$\pm$0.5 eV for Bi2212, and 3.3$\pm$0.5 eV for PLCCO). 
However, since the Pr $3d$ core level overlaps with the Cu $2p$ core level and also the Pr $4f$ valence band states overlap the Cu $3d$ states, we could not separate out the Cu $3d$ states from the Pr $4f$ states of PLCCO. Hence, we could not estimate $U_{d}$ for PLCCO, but instead we use the $U_{d}$ estimated for Bi2212. 
Next, using the estimated $U_{d}$ and $U_{p}$ values, and known values of $\Delta$ and $t_{pd}$, we obtain a set of parameter values for PLCCO and Bi2212 consistent with the experimental $J$ known from neutron or x-ray scattering using an optimization procedure\cite{Shesha}. The method also provides the effective one band parameters    $\tilde{U}$ and $\tilde{t}$ consistent with the experimental $J$. The results show that $\tilde{U}$/$\tilde{t}$ $\sim$9-10 for both PLCCO and Bi2212, and confirm the strongly correlated nature of the effective one-band singlet state in spite of the significantly different values of $U_{p}$.

\section{Experimental}

We have carried out XAS and Res-PES across the O K-edge of electron-doped Pr$_{1.3-x}$La$_{0.7}$Ce$_x$CuO$_{4}$  (PLCCO, with x = 0.1 ; $T_{c}$ = 27 K) and hole-doped  (Bi2212 ; $T_{c}$ = 90 K) to estimate $U_{p}$. For Bi2212, we also measured XAS and Res-PES across the Cu L-edge to estimate $U_{d}$. 
In addition, XAS and Res-PES across the O K-edge was measured for PLCCO with x = 0.0, which shows an antiferromagnetic metal ground state, to check the doping dependence of the two-hole Auger satellite.
The Bi2212 single crystal samples were prepared by the travelling solvent floating zone method as reported in the literature\cite{Li}, and characterized for their superconducting $T_{c}$ = 90 K. Res-PES across the O K-edge and Cu L-edge for Bi2212 was performed at BL17SU of SPring-8, Japan, with an energy resolution $\Delta$E = 0.2 eV. Bi2212 was peeled with scotch-tape in UHV and measured at T = 20 K. The Fermi level $E_F$ of
gold was measured to calibrate the energy scale. Low-energy off-resonant synchrotron valence band PES measurements ( h$\nu$ = 22.0 eV and 53.0 eV ) were carried out at BL21 of Taiwan Light Source, NSRRC, Taiwan.
The energy resolution was set to $\Delta$E =15 meV and the sample temperature was T = 10 K.
Single crystals of PLCCO with x = 0.0 and 0.10  were synthesized by the traveling-solvent floating-zone method and were protect annealed for 24 h at 800$^{\circ}$C\cite{Adachi}. The x = 0.1 composition showed a superconducting T$_C$ = 27 K.
XAS and Res-PES across the O K-edge for PLCCO was performed at BL2A of Photon Factory, Japan, with an energy resolution $\Delta$E = 0.2 eV. The XAS and Res-PES measurements were carried out at T = 200 K.
Low energy synchrotron PES with h$\nu$ = 16.5 eV and 55.0 eV
for PLCCO was performed at BL9A HiSOR and BL28A of Photon Factory, Japan, respectively. The energy resolution was set to $\Delta$E = 15 meV at HiSOR and at BL28A of Photon Factory. The measurements were carried out at T = 9 K and $E_F$ of
gold was measured to calibrate the energy scale.

\section{Results and Discussions}

\begin{figure}
\centering
\includegraphics[width=0.8\columnwidth]{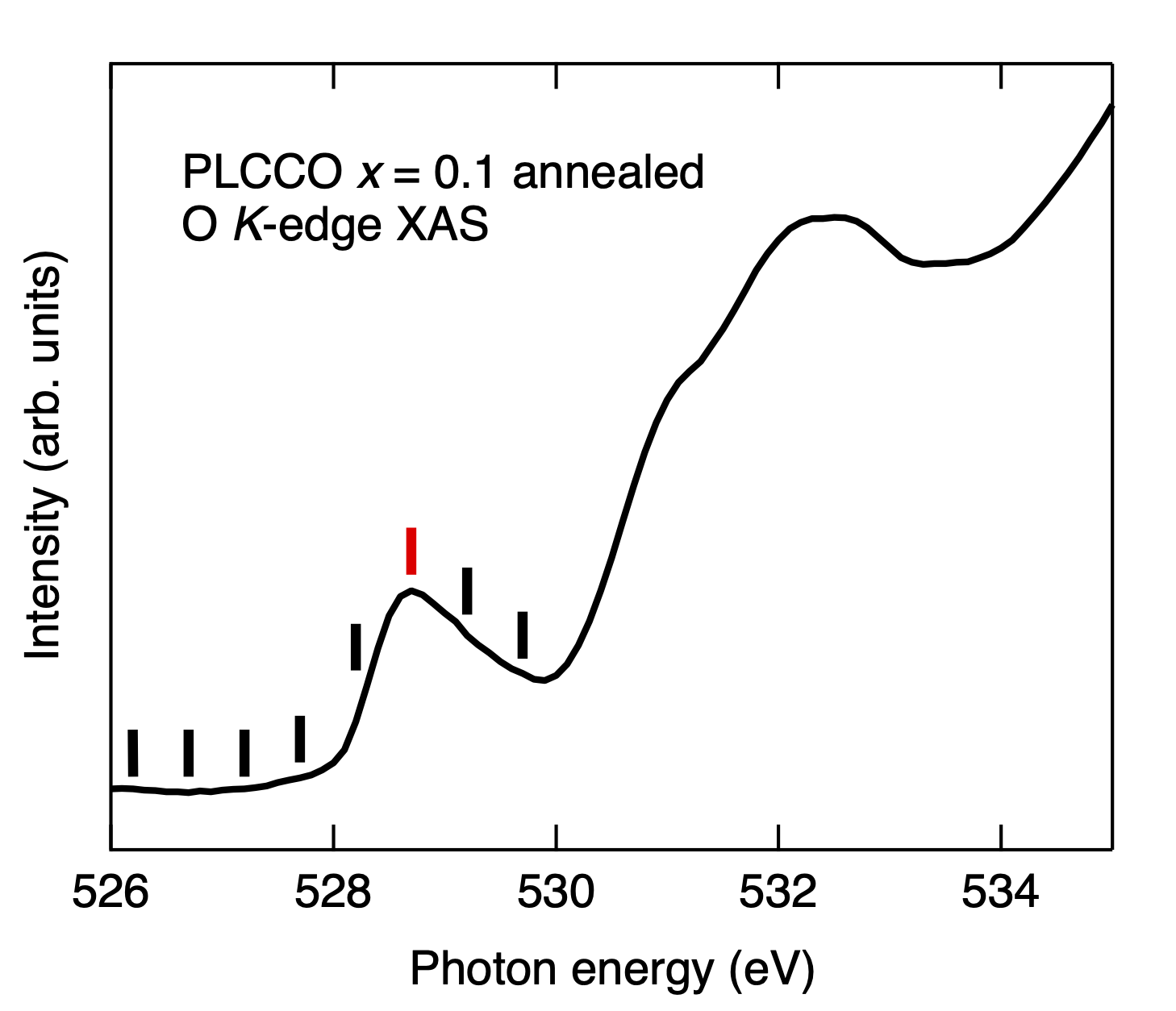}
\caption{\label{las} {The O K-edge (1s-2p) X-ray absorption spectrum of PLCCO, x = 0.1.}}
\end{figure}

\begin{figure}
\centering
\includegraphics[width=\columnwidth]{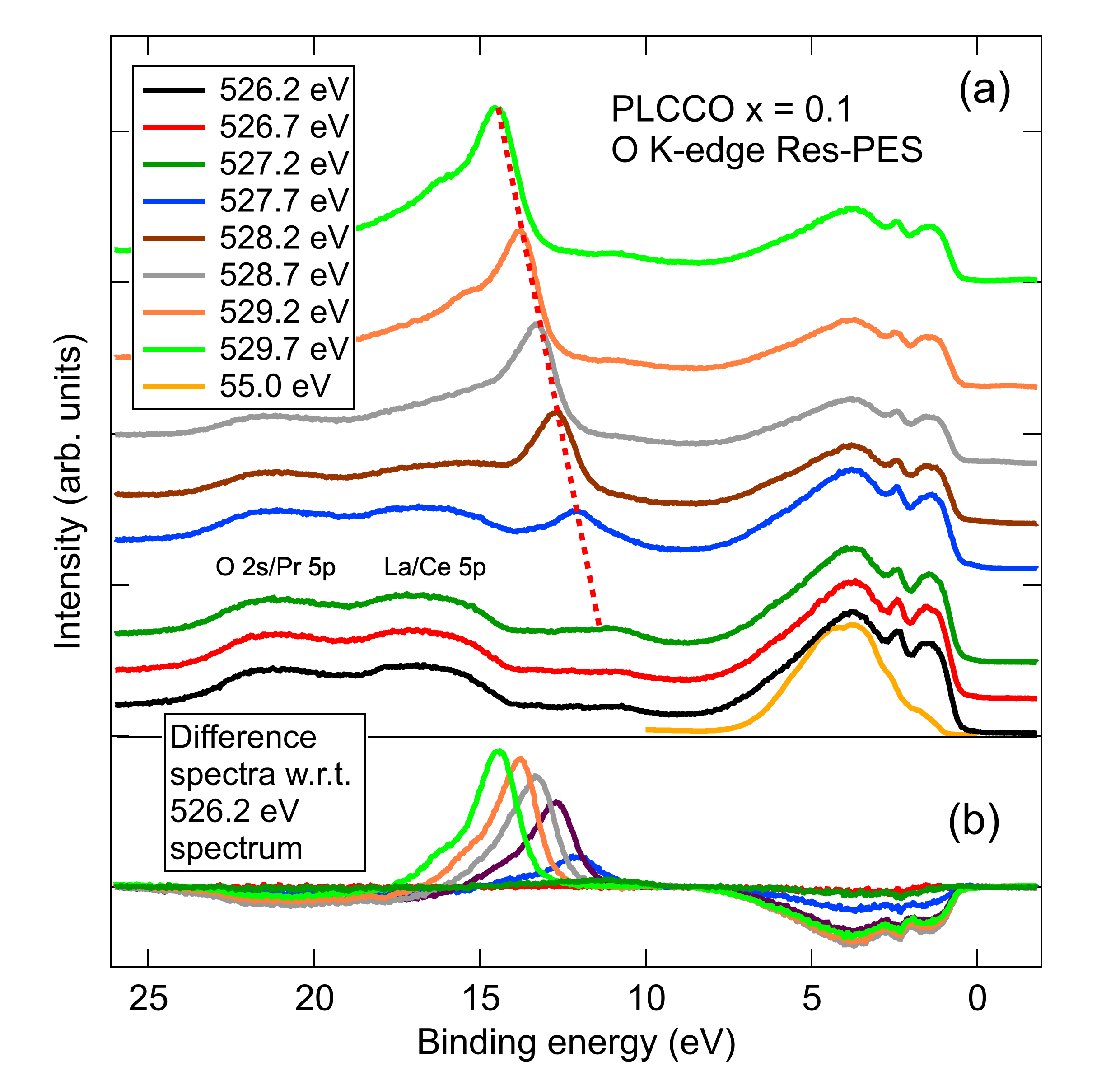}
\caption{\label{las} {(a) The  Res-PES spectra across the O K-edge (1s-2p) of PLCCO, x = 0.1, measured at photon energies marked with vertical bars in Fig. 2. The spectra are normalized at 8 eV BE. The off-resonance valence band spectrum measured with h$\nu$ = 55.0 eV is also shown. (b) The difference spectra obtained for higher energies with respect to the h$\nu$ = 526.2 eV spectrum.}}
\end{figure}

Figure 1 shows the O K-edge (1s-2p) XAS spectrum of PLCCO, x = 0.1, measured at T =  200 K over the incident photon energy range of h$\nu$ = 526-535 eV. It shows a small peak at $\sim$528.7 eV and a broad structure between 530-534 eV, with a weak shoulder at $\sim$531 eV. The states above 530 eV are attributed to the overlapping La, Ce and Pr 5d states hybridized with O 2p states\cite{Pellegrin}, while the 528-530 eV states are due to Cu 3d-O 2p hybridized states. The peak at 528.7 eV is quite similar to the lowest energy peak feature seen in the O K-edge XAS of electron-doped NCCO, which was analyzed as the unoccupied upper Hubbard band associated with Cu 3d states hybridizing with O p$_x$, p$_y$ states, while the p$_z$ states are mixed into the tail of the $\sim$531 eV shoulder\cite{Pellegrin}.

Figure 2(a) shows the O 1s-2p Res-PES spectra of PLCCO, x = 0.1, obtained using incident photon energies labelled by vertical tick marks in Fig. 1. The main valence band spectra shows three features consisting of a rounded peak at about 1.5 eV binding energy (BE), a small sharp feature at around 2.5 eV BE and a broad feature spread over 2.5-7.5 eV BE. The rounded peak is attributed to the mainly Pr$^{3+}$ occupied 4f$^{2}$ states which has a strong cross-section at these h$\nu$ values compared to Cu 3d states which are also expected over the same energies but hidden below the Pr 4f states. The small sharp feature at 2.5 eV BE is due to the Ce$^{3+}$ occupied 4f$^{1}$ states. This is confirmed by comparing the O 1s-2p Res-PES spectra of PLCCO, x= 0.0, which does not contain Ce, as discussed in Appendix A. The broad feature at 2.5-7.5 eV BE mainly consists of the O 2p states.
The valence band spectrum measured with  h$\nu$ = 55.0 eV is also shown in Fig. 2(a). It confirms the suppression of the Ce and Pr 4f states due to their low photo-ionization cross-sections at low incident h$\nu$, and also confirms the dominantly  O 2p PDOS character of the broad feature spread over 2.5-7.5 eV BE.

In this work, our main interest is to measure over higher binding energies and check for the O KVV Auger satellite feature which originates from a two-hole final state and provides a measure of $U_{p}$. As can be seen in Fig. 2(a), a weak feature seen at $\sim$11 eV BE shows a small increase in intensity on increasing the incident h$\nu$ from 526.2 to 527.2 eV. For higher h$\nu$ $>$ 527.2 eV, the feature gets strongly enhanced and shifts to higher BEs tracking the increase in h$\nu$ (red dashed line in Fig. 2(a)). This behavior is a signature of the Auger two-hole satellite. To characterize the evolution of the satellite, in Fig. 2(b), we have plotted the difference spectra with respect to h$\nu$ = 526.2 eV for all higher h$\nu$. The difference spectra shows a small intensity increase of the satellite feature at $\sim$11 eV BE for h$\nu$ = 527.2 eV (see Fig. 3 for an expanded y-scale figure). On increasing h$\nu$, it shows a systematic increase in intensity with an energy shift and a coupled suppression of the main O 2p valence band intensity.
The energy shift is seen with a small increase in intensity up to h$\nu$ = 529.7 eV, but a small increase of the main valence band intensity is also observed at h$\nu$ = 529.7 eV. The La and Ce 5p states are observed in Fig. 2(a) as weak bumps between $\sim$15-18 eV BE, while the Pr 5p states are between $\sim$ 20-23 eV and overlap with the O 2s states at $\sim$23 eV. A very similar behavior was observed in the O K-edge XAS and O 1s-2p Res-PES spectra of PLCCO, x = 0.0 (detailed in Appendix A), indicating a very similar O KVV Auger two-hole satellite.

\begin{figure}
\centering
\includegraphics[width=0.8\columnwidth]{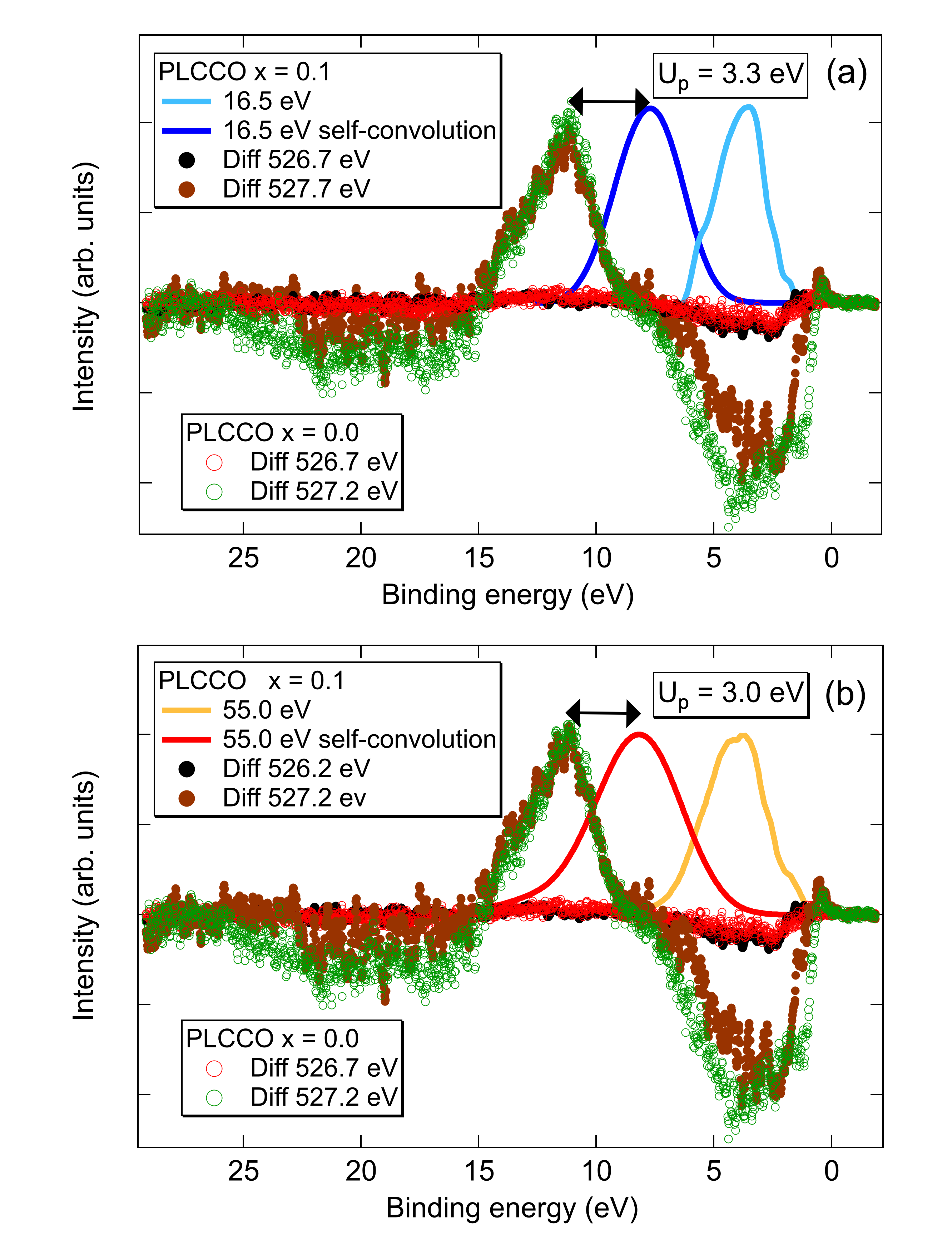}
\caption{\label{las} {(a) The valence band spectrum of PLCCO for x = 0.1 measured with h$\nu$ = 16.5 eV. From a numerical self-convolution of the one-hole valence band spectrum, we obtained the two-hole spectra. Comparing the two-hole spectrum with the difference spectra obtained at h$\nu$ = 527.2 eV (which shows the correlation satellite feature for x = 0.1), we could estimate $U_{p}$ = 3.3 eV$\pm$0.5 eV for h$\nu$ = 16.5 eV (b) Similarly, we could estimate $U_{p}$ = 3.0 eV$\pm$0.5 eV for h$\nu$ = 55 eV. The correlation satellite feature for x = 0.0 also lies at the same energy as for x = 0.1.}}
\end{figure}

In order to estimate $U_{p}$ using the Cini-Sawatzky method, we plot the PLCCO, x = 0.1 valence band spectra with h$\nu$ = 16.5 eV and 55.0 eV, as shown in Fig. 3(a) and 3(b), respectively. At these energies, the overall valence band spectrum is dominated by O 2p states but it can be seen that the spectrum with h$\nu$ = 55.0 eV is slightly broader than at h$\nu$ = 16.5 eV. From a numerical self-convolution of the one-hole valence band spectra, we obtained the two-hole spectra, also shown in Fig. 3(a) and 3(b). Comparing the two-hole spectra with the difference spectra of x = 0.0 and 0.1 obtained at h$\nu$ = 527.2 eV, which is the lowest energy that shows the two-hole correlation satellite feature, we estimate $U_{p}$ = 3.3 eV$\pm$0.5 eV for h$\nu$ = 16.5 eV and $U_{p}$ = 3.0 eV$\pm$0.5 eV for h$\nu$ = 55.0 eV. Thus, the estimated $U_{p}$ from the analyses using  h$\nu$ = 16.5 eV and 55.0 eV for x = 0.1 are quite close to each other.
Interestingly, as seen in Fig. 3, since the two-hole correlation satellite feature for x = 0.0 is observed at the same energy as for x = 0.1, it suggests that the strength of $U_p$ does not depend on the electron doping content.

\begin{figure}
\centering
\includegraphics[width=0.9\columnwidth]{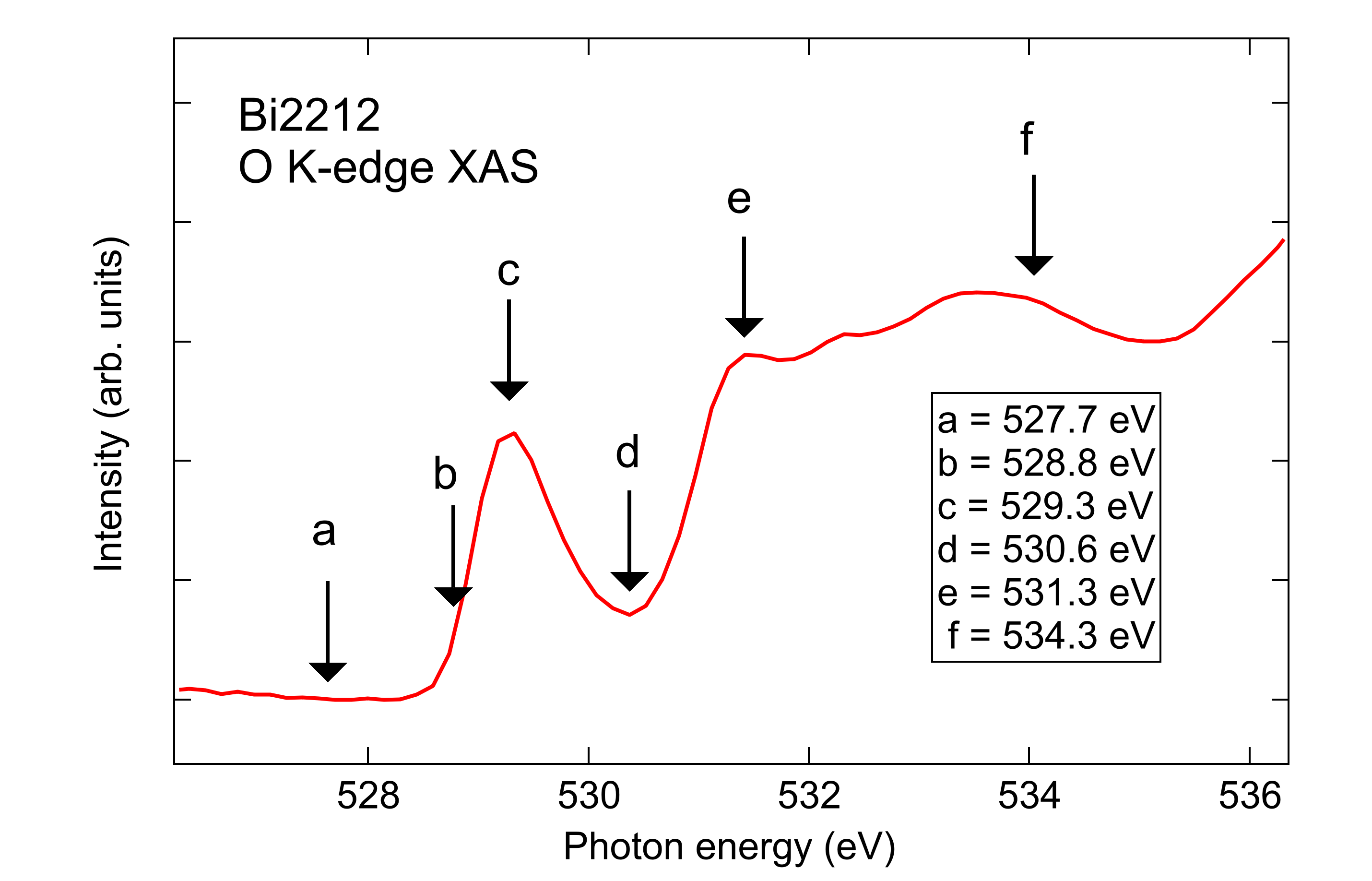}
\caption{The O K-edge (1s-2p) X-ray absorption spectrum of Bi2212. The photon energies labelled (a) - (f) were used to measure the Res-PES spectra across the O K-edge as discussed in Fig. 5.}
\end{figure}

\begin{figure}
\centering
\includegraphics[width=0.9\columnwidth]{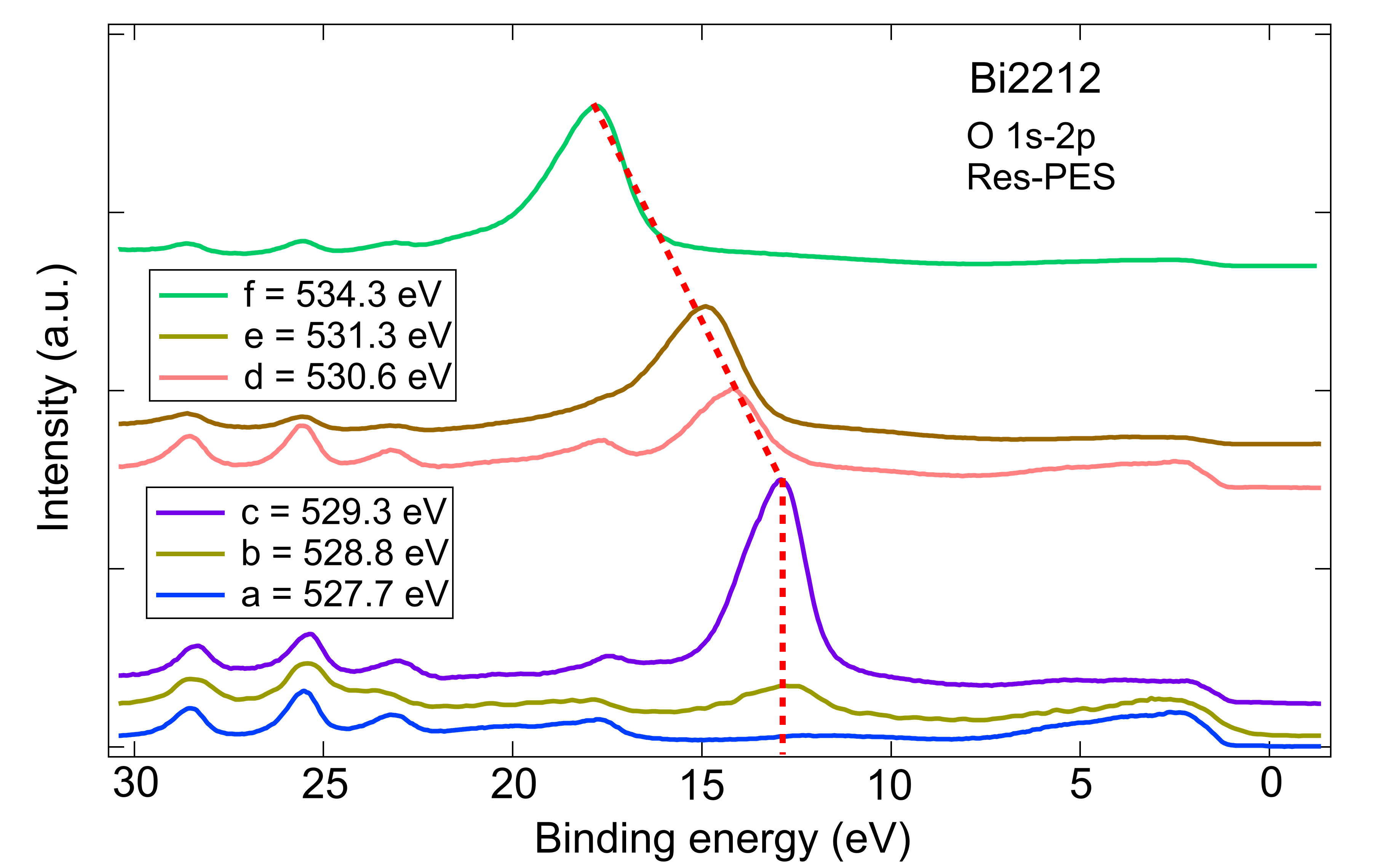}
\caption{The  Res-PES spectra measured across the O K-edge (1s-2p) of Bi2212 at photon energies labelled (a) - (f)  in Fig. 4. The spectra are normalized to the incident photon flux.}
\end{figure}

In Fig. 4, we plot the O K-edge (1s-2p) XAS spectrum of Bi2212 measured at T =  20 K over the incident photon energy range of h$\nu$ = 526-536 eV. The spectra are quite similar to early reports of the XAS of Bi2212\cite{Tjeng}. It shows a small peak at $\sim$528.8 eV, a shoulder at $\sim$531.35 eV which extends as a broad feature up to nearly 535 eV. The shoulder marks the onset of the upper Hubbard band associated with Cu 3d states bonding to O p$_x$, p$_y$ states while states above are attributed to the Bi, Sr and Ca states hybridized with O 2p states. It is well-known that the peak at 528.8 eV shows an intensity proportional to the doped hole states\cite{Himpsel,Nucker1}. Similar behavior was also seen in O K-edge XAS spectra of hole-doped La$_{2-x}$Sr$_{x}$CuO$_{4}$ \cite{CTChen}. At photon energies labelled (a)-(f), we then carried out O 1s-2p Res-PES spectra of Bi2212 to check for the two-hole Auger correlation satellite.

As shown in Fig. 5, the O 1s-2p Res-PES spectra of Bi2212, measured over a wide BE range of 30 eV, exhibits many shallow core levels, which are due to the Bi 5d, Ca 3p, O 2s and Sr 4p as labelled. The shallow core features between 17-30 eV BE allows us to consistently calibrate the on-resonance spectra in spite of the relatively weak intensities of the main valence band spectra between E$_F$ and about 7 eV BE.  Importantly, we see the peak feature at 12.8 eV BE systematically increase in intensity on increasing the incident h$\nu$ from 527.7 to 529.3 eV (a-c). At h$\nu$ = 530.6 eV, the intensity reduces, reflecting the dip in the XAS spectrum and then increases again for  h$\nu$ = 531.3 to 534.3 eV. From h$\nu$ = 529.3 eV to 534.3 eV, the feature systematically shifts to higher BEs tracking the increase in h$\nu$, confirming its Auger two-hole satellite character.

\begin{figure}
\centering
\includegraphics[width=0.9\columnwidth]{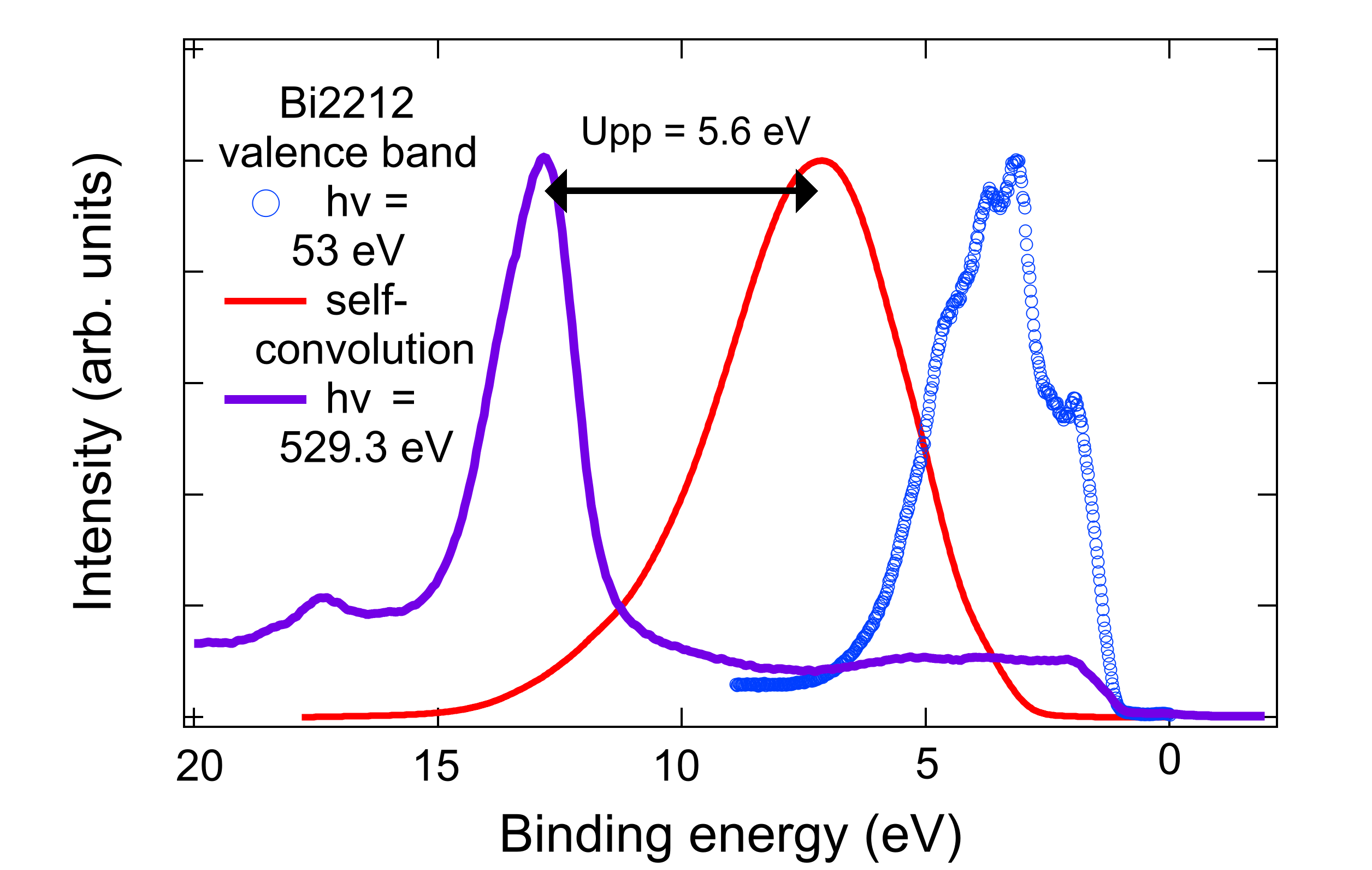}
\caption{The off-resonance valence band spectrum of Bi2212 measured with h$\nu$ = 53 eV, which represents the dominantly O 2p PDOS hybridized with Cu 3d states. The numerical self-convolution of the valence band spectrum is compared with the on-resonance  spectrum obtained with  h$\nu$ = 529.3 eV in order to estimate $U_p$. }
\end{figure}

In order to estimate $U_{p}$ for Bi2212, we measured the valence band spectrum with h$\nu$ = 53.0 eV, as plotted in Fig. 6. The spectrum shows the dominantly O 2p states hybridized with Cu 3d states, centered at about 3.5 eV BE, and very weak intensity with a step at the $E_F$. 
We have also measured the valence band with h$\nu$ = 22.0 eV (see Fig. 9, inset), but it is known that the Bi2212 spectrum shows relatively high intensity features due to the Bi-O derived O 2p states between 4-6 eV BE\cite{Eisaki}. Since we are interested in knowing the $U_{p}$ for the CuO$_2$-plane oxygen sites, we used the h$\nu$ = 53.0 eV spectrum to estimate $U_{p}$.  The numerical self-convolution of the one-hole h$\nu$ = 53.0 eV valence band spectrum is plotted together with the on-resonance spectrum obtained with h$\nu$ = 529.3 eV  spectrum, as shown in Fig. 6. The energy separation of the main peaks between these two spectra provides an estimate of $U_{p}$ = 5.6 eV$\pm$0.5 eV. Thus, the estimated $U_{p}$ = 5.6$\pm$0.5 eV for Bi2212 is larger than the value of $U_{p}$ = 3.3 eV$\pm$0.5 eV for PLCCO, and indicates that
 $U_{p}$ values can vary significantly for different families of cuprates.
While the origin of this difference in $U_{p}$ between PLCCO and Bi2212 is not clear, it is generally considered that the on-site Coulomb energy in a solid is strongly reduced from the atomic values due to solid-state screening. 
Considering the differences in the crystal structure of PLCCO and Bi2212, the smaller $U_p$ for PLCCO may be attributed to the generally smaller $\Delta$ (equivalently the smaller charge-transfer gap)  of the electron-doped cuprates compared to the hole-doped ones.

Next, we do the same exercise of estimating on-site Coulomb energy but for the Cu site, $U_{d}$, in Bi2212. 
Fig. 7 shows the Cu L-edge XAS spectrum which exhibits a typical single peak feature for the L$_3$ and L$_2$
edges. This is consistent with early work on Bi2212\cite{Nucker1, Bianconi}, which also reported polarization-dependent studies to characterize the Cu 3d states. It was shown that the single peak feature
was dominated by the 3d$_{x^2 - y^2}$ states, but also included about 15$\%$ 3d$_{z^2 - r^2}$ contribution \cite{Nucker1, Bianconi}. 
At photon energies labelled (a)-(h) marked in Fig. 7, we measured the Cu 2p-3d Res-PES spectra of BI2212 to check for the Cu two-hole Auger correlation satellite. Fig. 8 shows the valence band spectra measured over a wide energy range of 30 eV BE including the shallow core levels of Bi 5d, Ca 3p, O 2s and Sr 4p. The shallow core level positions help us to confirm the energy calibration. The spectral changes consist of a suppression or anti-resonance behavior of the main valence band, coupled to a large increase of the feature at about 12.5 eV BE. This peak shows a ten-fold increase in intensity on changing h$\nu$ from 930.6 eV to 933.4 eV corresponding to a resonant enhancement. Please note that the spectrum obtained with h$\nu$ = 933.4 eV is divided by a factor of 10. 
The spectrum with h$\nu$ = 933.4 eV is very similar to the early study by Brookes et al. which showed a strong resonant enhancement 
of the $\sim$12.5 eV satellite feature\cite{Brookes2001}. The authors further identified the feature at $\sim$12.5 eV as the atomic like $^1$G-state, the very weak feature at $\sim$16 eV as the $^1$S-state and the weak feature at $\sim$10 eV as the $^3$F-state.

\begin{figure}
\centering
\includegraphics[width=\columnwidth]{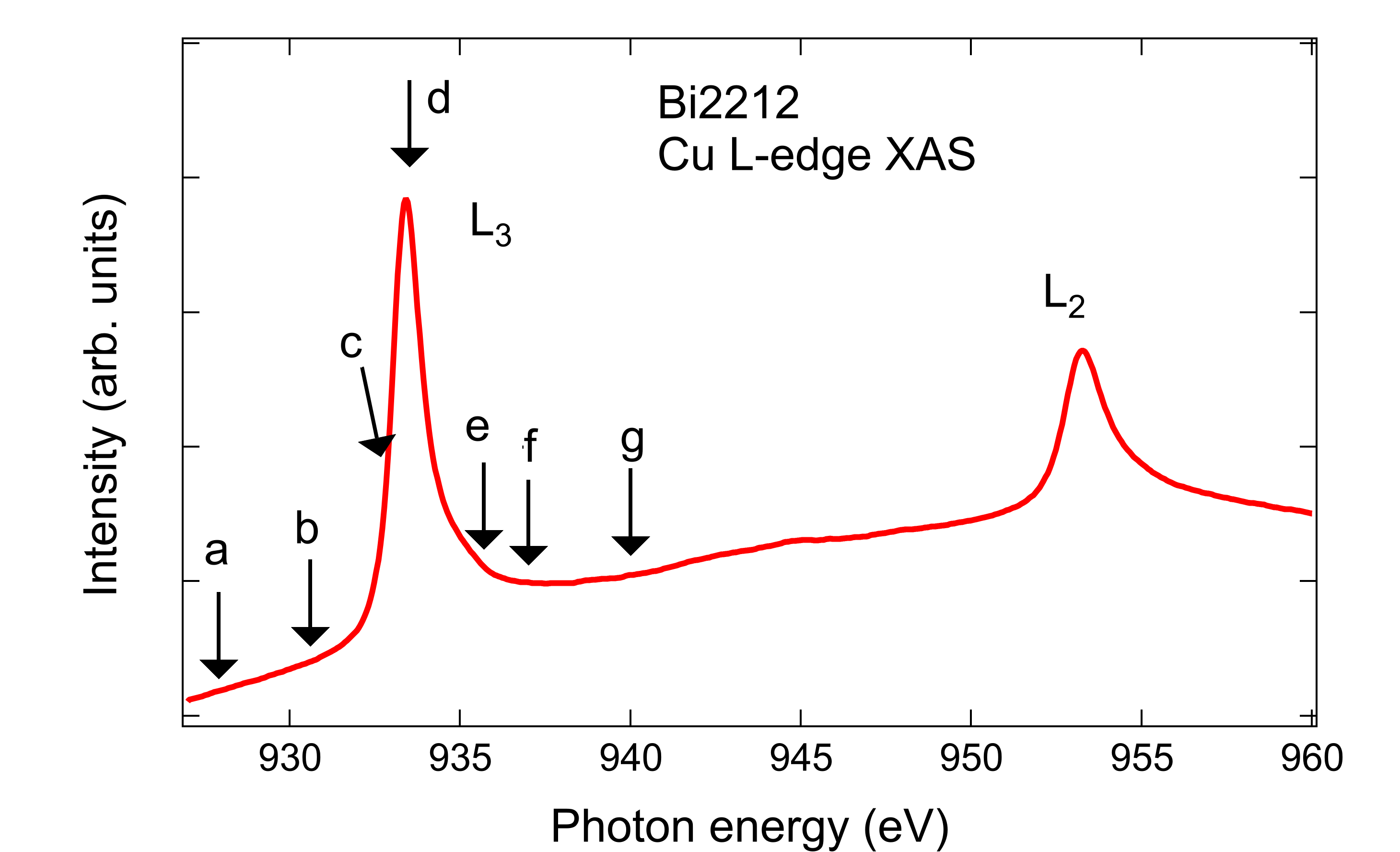}
\caption{(color online)  The Cu L-edge (2p-3d) X-ray absorption spectrum of Bi2212. The photon energies labelled (a)-(g) were used to measure the Res-PES spectra across the Cu L-edge, as discussed in Fig. 8.}
\end{figure}

\begin{figure}
\centering
\includegraphics[width=\columnwidth]{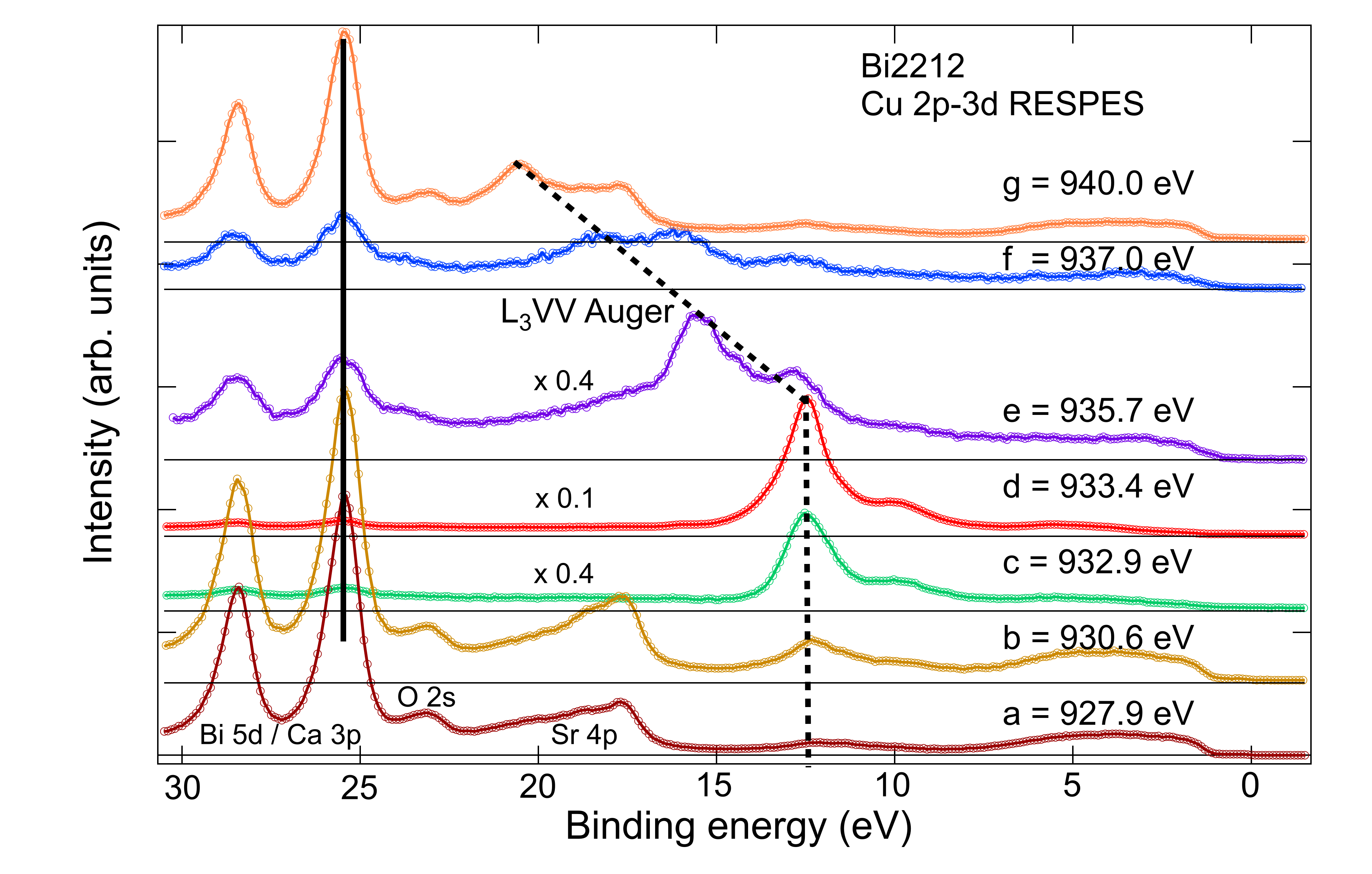}
\caption{The  Res-PES spectra measured across the Cu L-edge (2p-3d) of Bi2212 at photon energies labelled (a)-(g)  in Fig. 7. The spectra are normalized to the incident photon flux and in addition, for photon energies (c)-(e), the spectra were scaled by a factor to facilitate a comparative evolution of the L$_3$VV feature.}
\end{figure}

On increasing h$\nu$ further from 933.4 eV to 940.0 eV, the feature at 12.5 eV BE systematically moves to higher BE, tracking the increase in h$\nu$ and this indicates that the feature is the Cu L$_3$VV two-hole Auger satellite, consistent with early reports\cite{Tjeng}. In order to estimate $U_{d}$, we then measured the valence band of Bi2212 with h$\nu$ = 22.0 eV and compared it with the off-resonance spectrum obtained with h$\nu$ = 927.9 eV, as shown inset of Fig. 9. The h$\nu$ = 22.0 eV spectrum represents the valence band spectrum dominated by O 2p PDOS, which are hybridized with Bi and Cu valence band states. In particular, it was shown that the features between $\sim$4-8 eV BE are dominated by the Bi-O hybridized states which get suppressed even with h$\nu$ = 53.0 eV (see Fig. 6). Hence we used the h$\nu$ = 53.0 eV spectrum to estimate $U_p$ for the O 2p states associated with the CuO$_2$ planes. On the other hand, since the Cu 3d cross section dominates at h$\nu$ = 927.9, the h$\nu$ = 927.9 eV spectrum is considered to have an enhanced contribution of Cu 3d states, albeit hybridized with O 2p states.
In order to separate out the dominantly Cu 3d character PDOS, we normalized the spectra in the inset of Fig. 9 at 5.5 eV BE and obtained a difference spectrum, which is also plotted in the same inset. We then carried out a numerical self-convolution of the difference spectrum and compared it with the on-resonance h$\nu$ = 933.4 eV spectrum which showed the Cu L$_3$VV two-hole Auger satellite (Fig. 9, main panel). Although the numerical self-convolution shows weak features at BEs of 8 eV, 10 eV and 13 eV, we have checked that they arise from the structures between 4 and 6.5 eV BE in the difference spectra (inset, Fig. 9) associated with the Bi-O states lying at 4-8 eV BE. Hence, we used the main peak of the numerical self-convolution at 6 eV BE to get an estimate of average $U_d$ in Bi2212. The energy separation between the main peak of the numerical self-convolution and the main peak of the Cu L$_3$VV two-hole Auger satellite provides an estimate of $U_d$ = 6.5 eV$\pm$0.5 eV for Bi2212. Using the same method, a value of $U_d$ = 6.5 eV$\pm$0.5 eV was also estimated recently for the three layer cuprate superconductor HgBa$_2$Ca$_2$Cu$_3$O$_{8+\delta}$,\cite{AC2017} which shows the highest $T_{c}$ = 130 K at ambient pressure\cite{Yamamoto}. Having obtained estimates of $U_d$ and $U_p$, we applied it to determine the Heisenberg exchange $J$ and the relation between the effective one-band and three band Hubbard models for PLCCO and Bi2212.
But before that, we discuss below the very early work by deBoer et al.\cite{deBoer} which clarified the difference between the $U_d$ deduced from Auger spectra compared to the Hubbard $U_d$. 

For an atom M in a solid, the $U_d$ obtained from the two-hole Auger satellite is the energy cost for the ``reaction" $2(M^{+}) \rightarrow M + (M^{2+})$. Then, the value of $U_d$(Auger) is the difference between the first ionization energy ($I1$) and the second ionization energy ($I2$), i.e., $U_d$(Auger) = $I2 - I1$. However, the Hubbard $U_d$, is the energy cost for the``reaction" $2M \rightarrow (M^{-}) + (M^{+})$
i.e. Hubbard $U_d$ = $I1 - A$,  and corresponds to the difference between the first ionization energy I1 and the electron affinity $A$. While both the values represent the energy difference between one less electron and one more electron compared
to a reference state, the reference states $M$ and $M^{+}$ are obviously not the same. But the difference in the
estimated values of $U_d$(Auger) and Hubbard $U_d$ is expected to be small due to solid-state screening effects.\cite{deBoer}
It is noted that for the value of Hubbard $U_d$ for Cu, most of the 
literature uses values between 6-8 eV\cite{Fujimori,ZXS,Marel,Balzarotti,Ghijsen,Tjeng,BarDeroma,Johnston,
VeenendaalBi2212,Okada2009}, while we obtain $U_d$(Auger) = 6.5$\pm$0.5 eV, confirming that they are not very different. 

\begin{figure}
\centering
\includegraphics[width=\columnwidth]{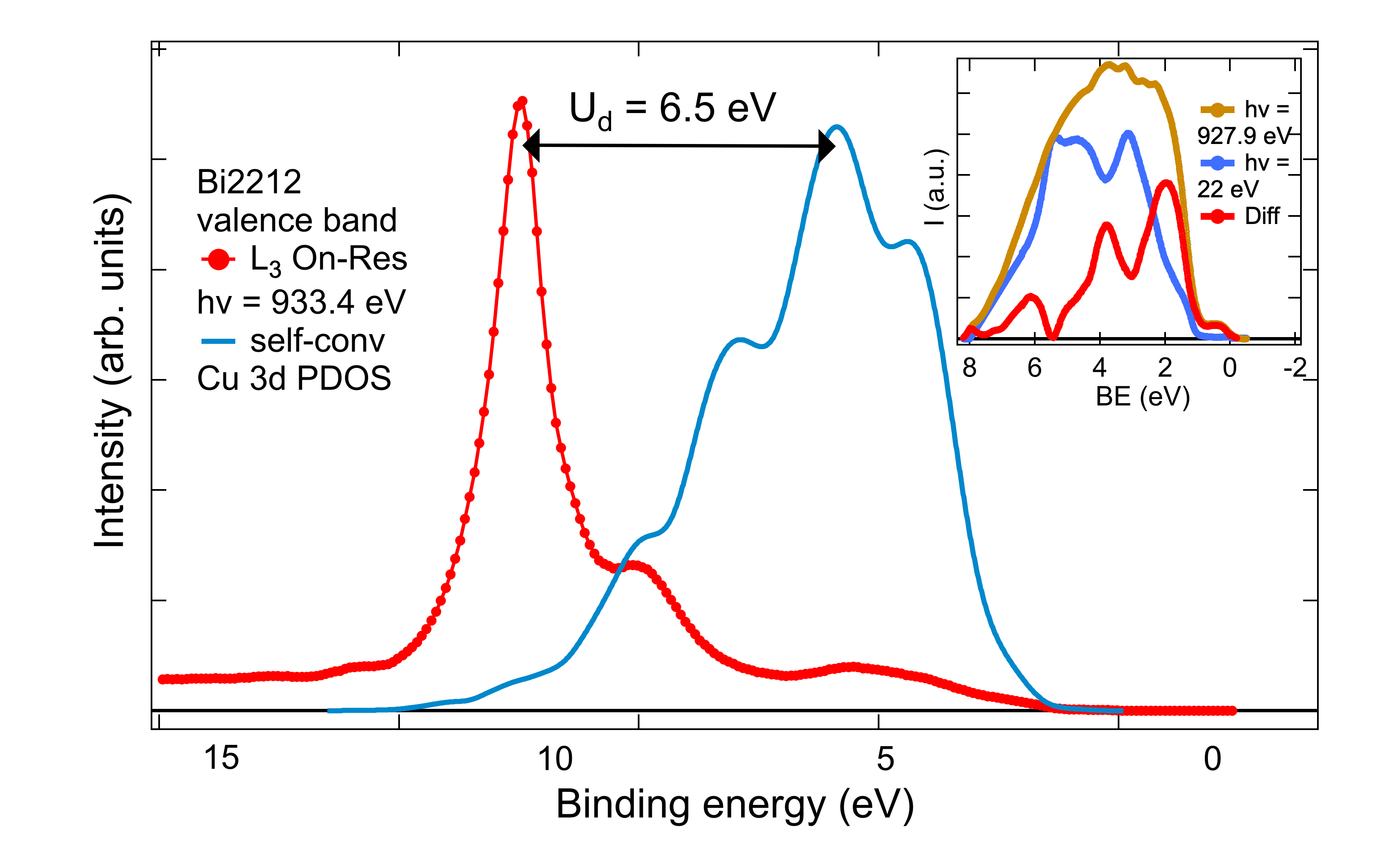}
\caption{The numerical self-convolution of the Cu 3d PDOS is compared with the on-resonance  spectrum obtained with  h$\nu$ = 933.4 eV in order to estimate $U_d$. Inset : The Cu 3d PDOS was obtained as the difference between the valence band spectrum of Bi2212 (h$\nu$ = 22.0 eV) and the off-resonance spectrum (h$\nu$ = 927.9 eV (Fig. 10).}
\end{figure}

\newpage

\onecolumngrid\

\begin{table}[t!]
	\begin{center}
	\caption{Electronic parameters ($\overline{U}_d$, $\overline{U}_p$, $\overline{t}$$_{pd}$, $\overline{\Delta}$ ) for cuprates from the three-band Hubbard model / cluster model calculations. The table also shows an optimized parameter set of ($t_{pd}$ and $\Delta$). $J$ is the nearest-neighbor Heisenberg exchange deduced from scattering experiments. See text for details.}
	
\begin{tabular}{|c|cccc|cc|c|ccc|}
\hline
&      &	&	& & ~~~~Optimized & set~~~~	&  &  & & \\
Compound&      $\overline{U}_d$	& $\overline{U}_p$  &	$\overline{t}$$_{pd}$	&	$\overline{\Delta}$	&$t_{pd}$ & $\Delta$~~~~& $J$(ref.no)		&~~~$\tilde{U}$~~~&~~~$\tilde{t}$~~~&~~~$\tilde{U}/\tilde{t}$~~~\\
(ref. no) 	&$\pm$0.5 eV	&$\pm$0.5 eV	 &$\pm$1.0 eV&$\pm$0.2 eV&eV &eV~~~~& meV&~~eV &~~eV &	\\
\hline

 Pr$_2$CuO$_4$(80)		&	8.0	&	4.1 	&	1.1	&	3.0	&	1.0	&	3.2~~~~	&	121 (86)	&~~3.16	&~~0.31	&	10.23	\\
&&&&&&&&&& \\
PLCCO(80) with				& 	6.5	&	3.3 	&	1.1	&	3.0   &	0.96	&	3.0~~~~	&  	121 (86)	&~~2.7	& ~~0.29 	& 	9.31 		\\
 exptal $U_d$, $U_p$ &&&&&&&&&& \\
\hline

Bi2212(78)				&	8.5	&	4.1	&	1.13	&	3.2	&	1.1	&	3.5~~~~	&	161 (87)	&~~3.34	&~~0.37	&	9.31		\\
 Bi2212(79)				&	7.7	&	6.0	&	1.5	&	3.5	&	1.2	&	3.7~~~~	&	161 (87)	&~~3.59	&~~0.38	&	9.44		\\
 &&&&&&&&&& \\
 Bi2212(79) with 				& 	6.5	& 	5.6 	&	1.5 	&	3.5	& 	1.16	&	3.5~~~~	& 	161 (87)	& ~~3.2	&~~0.36	& 8.9 			\\
 exptal $U_d$, $U_p$ &&&&&&&&&& \\
\hline

\end{tabular}
	\end{center}

\end{table} 
\twocolumngrid\

In a recent study \cite{Shesha}, we have developed an optimization procedure to estimate effective one-band Hubbard model parameters $\tilde{U}$ and $\tilde{t}$ using reported three-band parameters $t_{pd}, \Delta, U_d$ and $U_p$ from theoretical studies\cite{Johnston} as well as cluster model calculations \cite{VeenendaalBi2212,Okada2009}. In this procedure, the Heisenberg exchange $J$ calculated  using a downfolding method \cite{Koch} for a Cu$_2$O cluster model employing the three-band Hamiltonian in the hole picture, is given by
\begin{equation}
\label{eq:khomskiij}
J = 4\frac{t_{pd}^4}{\Delta^2} \left[ \frac{1}{U_d} + \frac{1}{\Delta + U_p/2} \right].
\end{equation}
This agrees with the expression obtained by fourth-order perturbation theory \cite{zs1987,EskesJefferson,Khomskii,jbs2020}, but in the approximation of inter-site Coulomb interaction $U_{pd}$ = 0 (which is typically smaller than $U_p$ and $U_d$\cite{Werner}) and the oxygen-oxygen hopping $t_{pp}$ = 0 (since we used a Cu$_2$O cluster). If we now write $J = 4\tilde{t}^2/\tilde{U}$, then we can identify
\begin{equation}
\label{eq:ueff}
\tilde{t} = \frac{t_{pd}^2}{\Delta}, ~~~~\frac{1}{\tilde{U}} = \frac{1}{U_d} + \frac{1}{\Delta + U_p/2}.
\end{equation}
As explained in ref. 58, this expression for $J$ does not lead to a satisfactory agreement with $J$ reported from neutron and x-ray scattering measurements\cite{Nd2CuO4,Wang}, if we directly use reported values of three-band parameters. Using an optimization procedure, we first find values which provide a good agreement with $J$ known from scattering experiments.  It was found that the energy cost was minimal for the second optimization procedure (described in ref. 58), in which we modify the parameters ($\bar{t}$$_{pd}$, $\bar{\Delta}$ ; columns 3,4 in Table 1) and obtain optimal values ($t_{pd}$ and $\Delta$ ; columns 5,6 in Table 1). The optimal values are sufficiently close to values of $\bar{t}$$_{pd}$, $\bar{\Delta}$ using the three-band model or cluster model calculations reported in the literature. Next, we use our measurements of $U_{d}$ and $U_{p}$, and optimal values $t_{pd}$, $\Delta$  to estimate the one-band parameters $\tilde{U}$ and $\tilde{t}$. The results are summarized in the Table 1. The results show that $\tilde{U}$/$\tilde{t}$ $\sim$9 - 10 for both PLCCO and Bi2212, and this confirms the strongly correlated nature of the effective one-band singlet state.\cite{Shesha}
It is very interesting to note that the estimated one-band parameters $\tilde{U}$ and $\tilde{t}$ show small differences for PLCCO and Bi2212, although the $U_p$ values are significantly different for them. If one looks at the small differences between PLCCO and Bi2212 more closely, one can see that the smaller $t_{pd}$ for PLCCO (due to its longer in-plane lattice parameter) is responsible for
 the smaller $J$ and $\tilde{t}$, in spite of the smaller $\Delta$ and $U_p$. On the other hand, the smaller $\Delta$ and $U_p$ do play a major role in reducing $\tilde{U}$ in PLCCO. In contrast, the larger $t_{pd}$ for Bi2212 (due to its shorter in-plane lattice parameter) is responsible for the larger $J$ and $\tilde{t}$. Although a larger $\Delta$ and $U_p$ results in a relative increase in $\tilde{U}$ for Bi2212, the net result is still a larger $J$ for Bi2212 compared to PLCCO.
 
More interestingly, the obtained values of $\tilde{t}$ = 0.29 eV (for PLCCO) and $\tilde{t}$ = 0.36 eV (for Bi2212) are quite close to the values of the primary or nearest-neighbor (NN) hopping $t$ = 0.26 eV (for PLCCO) and 0.36 eV (for Bi2212) estimated from fitting the ARPES Fermi surfaces of PLCCO\cite{Horio_Thesis} and Bi2212\cite{Valla2}. It is noted that the tight binding fits for PLCCO and Bi2212 also employed a second NN hopping $t'$ ( = 0.24$t$ and 0.3$t$, respectively ) and for Bi2212, an additional out-of-plane hopping $t_{\perp}$ ( = 0.3$t$ ), which are relatively small. Similar results have been reported for La$_2$CuO$_4$
and Sr$_2$CuO$_2$Cl$_2$ for which the neutron scattering results could be explained by an effective extended one-band model. For La$_2$CuO$_4$, a dominant NN hopping  $t$ = 0.33 eV implied an effective $U/t$  = 8.8 with $U$ = 2.9 eV, but in addition to the NN exchange $J$ = 138 meV, it was important to include a ring exchange term with $J_c$ = 38 meV, and the second NN and third NN exchange $J' = J^{''} = 2$ meV\cite{Coldea}. For Sr$_2$CuO$_2$Cl$_2$, the authors used a
$t-t'-t^{''}-J$ model and obtained $t$ = 0.35 eV, $t' = 0.12$ eV, $t^{''} = 0.08$ eV, and with a $J = 0.14$ eV \cite{Leung,Kim}, it implied an effective $U/t$ = 10 with $U$ = 3.5 eV. All these cases suggest that the NN hopping
$t$ and $U$ can be considered to be $\tilde{t}$ and $\tilde{U}$ of the effective one-band model.

Thus, in spite of the differences in PLCCO and Bi2212, $\tilde{t}$ plays an important role in determining the value of $J$ and also results in a very similar value of $\tilde{U}$/$\tilde{t}$ $\sim$9-10. 
Several studies have  emphasized $J$ as being one of the most important parameters to achieve high-temperature superconductivity exhibited by the family of cuprates \cite{Keimer,Sr2CuO3, Nd2CuO4,YBCO,Dean, Bi2212,Wang,Peng,Lipscombe,Braicovich,Dean2,Levy}.
It is clear from eqn. (1) that $U_{p}$,  $U_{d}$,  $\Delta$ and $t_{pd}$, all play an important role in determining the Heisenberg exchange $J$. Finally, using eqn (2) and
writing $J = 4\tilde{t}^2/\tilde{U}$ in the effective one band Hubbard model form provides a bridge to understand the connection between the effective one-band and three band Hubbard models of the cuprates\cite{Aligia01,Aligia02}. While they have been often considered as distinct models, but in essence, as the present results show, they are truly equivalent.

\section{Conclusions}

In conclusion, the Cini-Sawatzky method was employed to obtain the experimental values of $U_{d}$ ( = 6.5$\pm$0.5 eV) for Bi2212 and $U_{p}$ for
 for Bi2212 ( = 5.6$\pm$0.5 eV) and PLCCO ( = 3.3$\pm$0.5 eV). This indicates that the
 $U_{p}$ values can vary significantly in different families of cuprates.
Using the estimated $U_{d}$ and $U_{p}$ values, and known values of $\Delta$ and $t_{pd}$, we could obtain a set of optimal parameter values for PLCCO and Bi2212 consistent with the experimental $J$ known from neutron or x-ray scattering. We also obtained the effective one band parameters $\tilde{U}$ and $\tilde{t}$ for the experimental $J$. The results show that $\tilde{U}$/$\tilde{t}$ $\sim$9-10 for both PLCCO and Bi2212, and confirm the strongly correlated nature of the effective one-band singlet state.

\section{Acknowledgements} 
The synchrotron radiation experiments were performed at : BL17SU, SPring-8, Japan with the approval of RIKEN (Proposal No. 20140019) ; BL 2A and BL 28 Photon Factory, Japan (2014G177, 2012G075, 2012S2-001); BL9A HiSOR, Japan; BL 21A Taiwan Light Source, NSRRC, Taiwan. We thank H. Anzai, M. Arita, K. Ono, H. Suzuki, K. Koshiishi and D. Ootsuki for valuable technical support. This work was supported by JSPS KAKENHI (Grant Numbers JP19K03741, JP22K03535 and JP19H01841) and by the ``Program for Promoting Researches on the Supercomputer Fugaku" (Basic Science for Emergence and Functionality in Quantum Matter, JPMXP1020200104) from MEXT. AC thanks the National Science and Technology Council (NSTC) of the Republic of China, Taiwan for financially supporting this research under Contract No. MOST 111-2112-M-213-031. 

\section{Appendix A : XAS and Res-PES of PLCCO, x = 0.0}

\begin{figure}
\centering
\includegraphics[width=0.78\columnwidth]{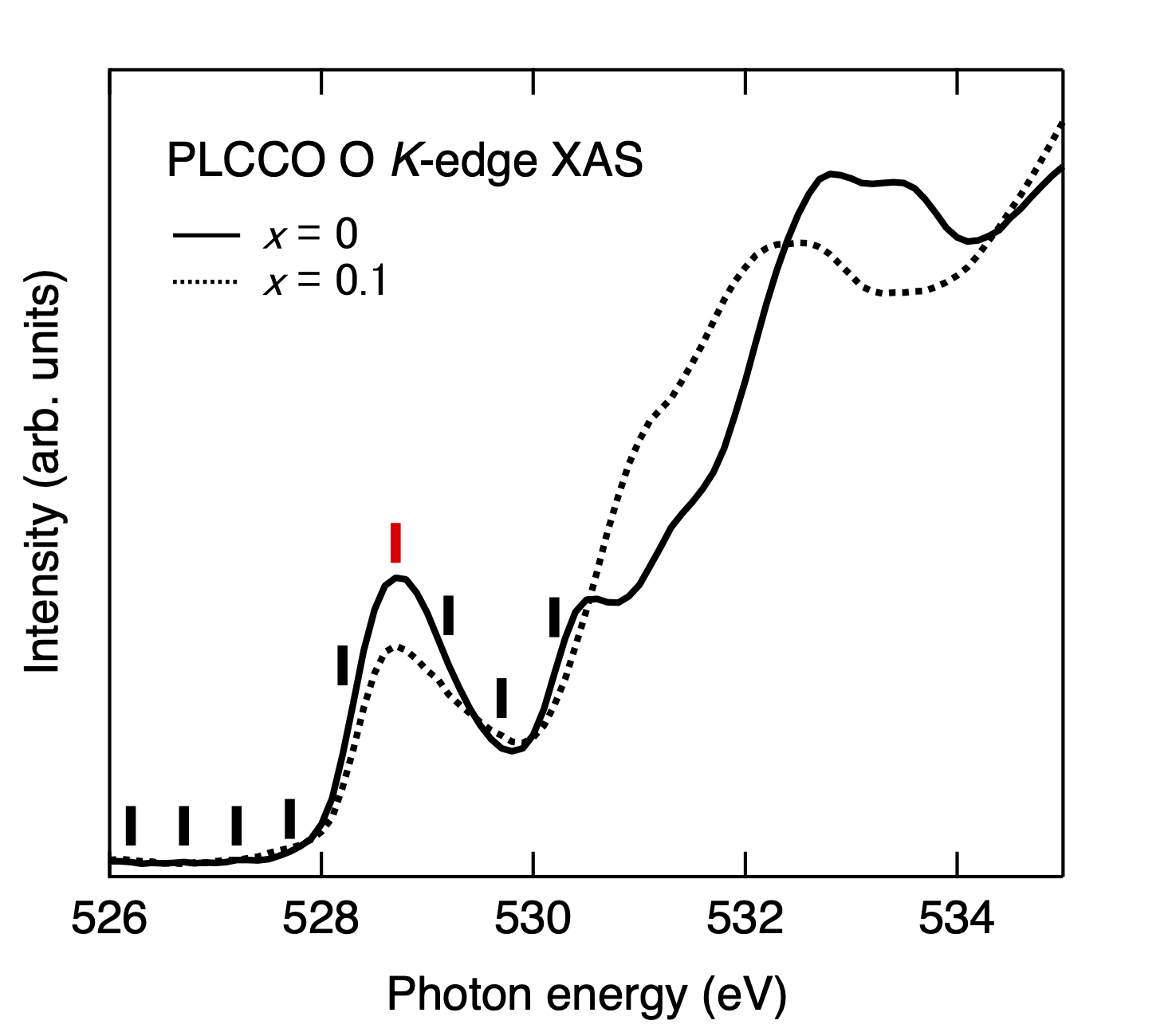}
\caption{\label{las} {Comparison of the O K-edge (1s-2p) X-ray absorption spectra of PLCCO, x = 0.0 and 0.1.}}
\end{figure}

\begin{figure}
\centering
\includegraphics[width=\columnwidth]{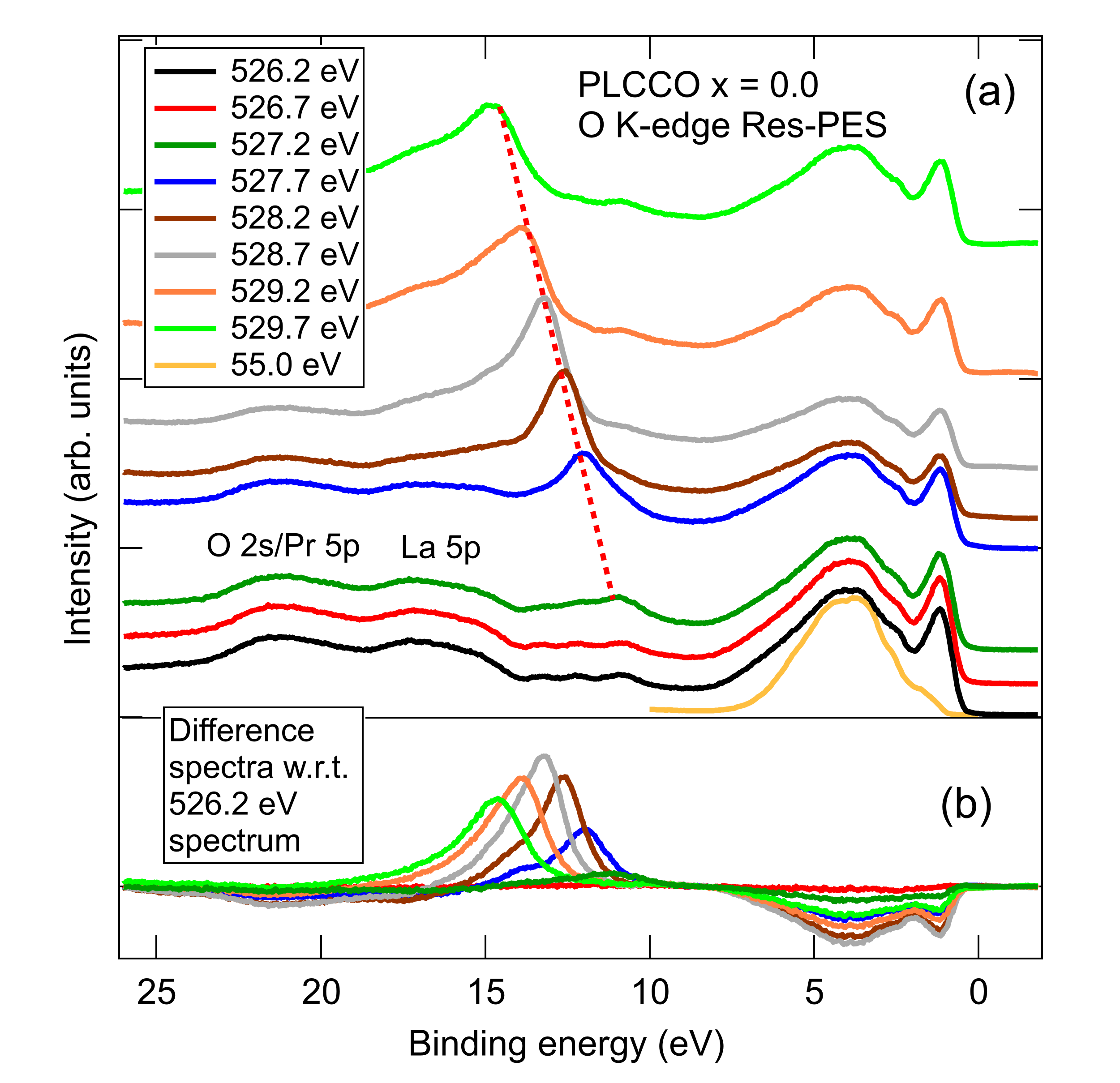}
\caption{\label{las} {(a) The  Res-PES spectra across the O K-edge (1s-2p) of PLCCO, x = 0.0, measured at photon energies marked with vertical bars in Fig. 1. The spectra are normalized at 8 eV BE. The off-resonance valence band spectrum measured with h$\nu$ = 55 eV for x = 0.1 is also shown for comparison. (b) The difference spectra for higher energies obtained with respect to the h$\nu$ = 526.2 eV spectrum.}}
\end{figure}

Figure 10 shows the O K-edge (1s-2p) XAS spectrum of PLCCO, x = 0.0, measured at T =  200 K over the incident photon energy range of h$\nu$ = 526 - 535 eV. It shows a small peak at $\sim$528.7 eV, a weak shoulder at $\sim$530.5 eV and a broad structure at 532 - 535 eV. The high energy states above 532 eV are attributed to the La and Pr 5d states hybridized with O 2p states\cite{Pellegrin}, while the 528 - 530 eV states are due to Cu 3d- O 2p hybridized states. The peak at 528.7 eV is also quite similar to the lowest energy peak feature seen in the O K-edge XAS of electron-doped NCCO, which was analyzed as the unoccupied upper Hubbard band associated with Cu 3d states hybridizing with O p$_x$, p$_y$ states, while the shoulder at $\sim$530.5 eV is due p$_z$ states\cite{Pellegrin}. Comparing the x = 0.0 and x = 0.1 spectra as shown in Fig. 10, the small peak associated with the upper Hubbard band at 528.7 eV shows relatively lower intensity in x = 0.1 compared to x = 0.0. This confirms the higher electron doping content in x = 0.1 with respect to x = 0.0. 

The O 1s-2p Res-PES spectra of PLCCO, x = 0.0 shown in Fig. 11(a) are quite similar to that of x = 0.1 shown in Fig. 2(a). There are small differences such as the small Ce$^{3+}$ peak at around 2.5 eV BE is missing in x = 0.0 and the mainly Pr$^{3+}$ occupied 4f$^{2}$ states at 1.5 eV BE is sharper with slightly higher intensity.  The Res-PES spectra also show the two-hole Auger satellite feature at $\sim$11 eV BE, which shifts to higher BEs tracking the increase in 
h$\nu$ (red dashed line in Fig. 11(a)). The resonance behavior of the satellite was confirmed by plotting the difference spectra with respect to h$\nu$ = 526.2 eV, as shown in Fig. 11(b). The satellite starts getting enhanced at h$\nu$ = 526.2 eV, and its energy position and spectral shape is very similar to the satellite observed for x = 0.1, as shown in Fig. 3. For higher h$\nu$, the difference spectra show an increase of the satellite intensity and shift to higher BE, coupled with a suppression of the main valence band states till 
h$\nu$ = 528.7 eV. This is followed by a suppression of the satellite coupled with a recovery of the main valence band states at h$\nu$ = 529.7 eV. The La 5p states are observed in Fig. 11 (a) as weak features between $\sim$15-18 eV BE, while the Pr 5p states occur between $\sim$20-23 eV and overlap with the O 2s states at $\sim$23 eV.
The valence band spectrum measured with  h$\nu$ = 55.0 eV for x = 0.1 is also shown in Fig. 11(a). It shows that the broad O 2p states spread over 2.5-7.5 eV BE for x = 0.0 with higher h$\nu$ are quite similar to the O 2p states for x = 0.1.
It is noted that although we did not measure the low energy h$\nu$ = 16.5 eV or 55.0 eV valence band spectra to estimate $U_p$ for x = 0.0, the BE shifts of the La $3d$, Pr $3d$, and O $1s$ core-level peaks were measured by x-ray photoemission spectroscopy\cite{Horio_Thesis}. The results indicated a chemical potential shift of $<$0.3 eV from x = 0.0 to x = 0.1. Since the O 2p feature between 2.5-7.5 eV BE for x = 0.1 matches closely with the O 2p feature for the x = 0.0 spectra measured with higher h$\nu$, it indicates that for x = 0.0, the shift in the O 2p  PDOS in the valence band is also $<$0.3 eV. Accordingly, the change in $U_p$ for x = 0.0 is considered to be within the error bar ($\pm$0.5 eV) of the $U_p$ estimated for x = 0.1.

\end{document}